\documentclass[superscriptaddress,aps,prb,twocolumn]{revtex4-1}
\usepackage{graphicx} 
\usepackage{bm} 
\usepackage{color}
\usepackage{amsmath}

\newcommand{\sig}{ \mbox{\boldmath{$\sigma$}}}

\newcommand{\Del}{ \mbox{\boldmath{$\Delta$}}}

\begin{document}

\title{Exchange intervalley  scattering and magnetic phase diagram of transition metal dichalcogenide monolayers}

\date{\today}

\author{Dmitry Miserev, Jelena Klinovaja, and Daniel Loss} 
\affiliation{Department of Physics, University of Basel, Klingelbergstrasse 82, CH-4056 Basel, Switzerland}

\begin{abstract}
We analyze magnetic phases of monolayers of transition metal dichalcogenides that are two-valley materials with electron-electron interactions. The exchange inter-valley scattering makes two-valley systems less stable to the spin fluctuations but more stable to the valley fluctuations. We predict a first order ferromagnetic phase transition governed by the non-analytic and negative cubic term in the free energy that results in a large spontaneous spin magnetization. Finite spin-orbit interaction leads to the out-of-plane Ising order of the ferromagnetic phase. Our theoretical prediction is consistent with the recent experiment  on electron-doped monolayers of MoS$_2$ reported by Roch {\em et al.}~\cite{roch}
The proposed first order phase transition 
can also be tested by measuring the linear magnetic field dependence of the  
spin susceptibility in the paramagnetic phase 
which is a direct consequence of the
non-analyticity of the free energy.
\end{abstract}

\maketitle

\section{Introduction}

Multi-valley materials provide an alternative to spintronics where, instead of spin projections, multiple valleys are used as pure states for quantum bits. This additional valley degree of freedom also provides the ground for complex phase diagrams for multi-valley materials \cite{kelly,bloss}. Transition metal dichalcogenides (TMDs) are particularly promising two-valley materials for valleytronics \cite{wang} due to the large spin-orbit interaction (SOI) \cite{zhu,xiao,chei,kormanyos,kormanyos2013,klinovaja,review_1,pisoni} and the control of the valley polarization by external magnetic fields \cite{kormanyos,cheng,aivazian}. Large SOI also results in the valley-dependent light absorption which yields another way to control the valley polarization at low densities \cite{zeng,mak}. 

Effects of electron-electron interactions in TMDs were considered previously in Refs.~\cite{donck,mukherjee,braz}
The results predicted in Refs.~\cite{donck,mukherjee} are valid if the exchange inter-valley (EIV) interaction is negligible. The limit considered in Ref.~\cite{braz} corresponds to a SOI strength which is much larger than the Fermi energy. In this limit the EIV scattering is forbidden due to the spin-valley locking. Signatures of the valley ferromagnetic phase predicted in Ref.~\cite{braz} were observed in WS$_2$ \cite{scrace} where the SOI gap is relatively large $\sim 40 \,$meV.

In this paper we consider the opposite limit when the SOI gap is small compared to the Fermi energy. This scenario can be realized in the conductance band of MoS$_2$ where the SOI gap is on the order of a few meV. We predict a first-order ferromagnetic phase transition governed by non-analytic terms in the free energy. 
Such terms for one-valley  materials have been calculated in Refs.~\cite{maslovsaha,maslovchubukov} They originate from the infrared electron-hole fluctuations near the Fermi surface in two-dimensional (2D) interacting systems \cite{chubukov}. The one-valley calculations \cite{maslovsaha,maslovchubukov,chubukov,zakmaslov,zakQ}, however, cannot be applied to two-valley materials if the EIV scattering must be taken into account.
Here we argue that the EIV scattering is important as it favors a ferromagnetic instability over valley and spin-valley instabilities. We also argue that the SOI breaks the $O(3)$ symmetry of the ferromagnetic phase leading to the Ising order.
Similar scenario is realized in one-valley materials with the Rashba SOI \cite{zakmaslov,zakQ}.
This work is motivated by the recent experiment in $n$-doped monolayer of MoS$_2$ \cite{roch} where the ferromagnetic Ising order has been detected. 
 The non-analytic cubic term in the free energy can be directly probed as it results in an unusual linear magnetic field dependence of the spin susceptibility in the paramagnetic phase.

\begin{figure}[t]
	\includegraphics[width=0.9\columnwidth]{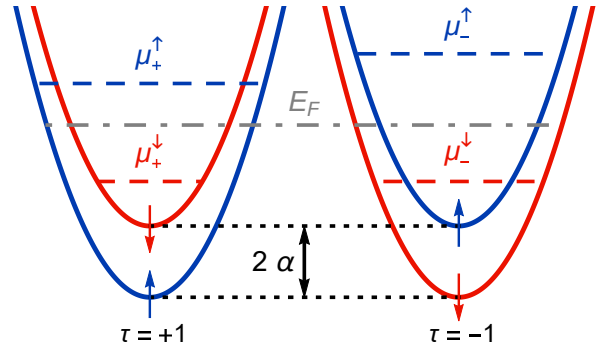}
	\caption{Electron spectrum of monolayers of TMDs described by the Hamiltonian defined in Eq.~(\ref{ham}). Blue (red) bands correspond to  spin up (down) along the $z$-axis, marked by up (down) arrows, $2 \alpha$ is the SOI gap. Effect of the electron-electron interaction is schematically illustrated by the chemical potentials $\mu^\uparrow_\pm$ and $\mu_\pm^\downarrow$ which are different from the Fermi level $E_F$ of the non-interacting 2DEG (dashed-dotted).}
	\label{fig:spec}
\end{figure}

Our paper is organized as follows. In Section II we introduce the matrix elements of the electron-electron interaction. In Section III we derive that the EIV interaction makes  possible only ferromagnetic and paramagnetic phases, other phases cannot develop due to the stabilizing effect of the EIV interaction on the valley and spin-valley fluctuations. In Section IV we outline the Ginzburg-Landau theory with the negative cubic non-analyticity in case of a single order parameter. In Section V we present the non-analytic cubic correction to the free energy of two-valley 2D electron gas.  Conclusions are given in Section VI. The diagrammatic calculations and other technical details are deferred to Appendices A-C.

\section{Electron-electron interaction}

Monolayers of TMDs are two-valley materials.
The electron spectrum of such materials can be effectively described by the following free-particle Hamiltonian \cite{kormanyos}:
\begin{equation}
H = \frac{k^2}{2 m} - \alpha \sigma_z \tau,
\label{ham}
\end{equation}
where $\bm k = (k_x, k_y)$ is the 2D momentum, $m$ is the effective mass, $2 \alpha$ is the SOI gap, $\sigma_z$ is the spin Pauli matrix, the index $\tau = \pm 1$ labels the valley. The spectrum of $H$ consists of four bands, see Fig.~\ref{fig:spec}.  Each pair of bands is split by the SOI $\alpha$ which also lifts the spin degeneracy and, hence, defines the spin quantization axis.

\begin{figure}[t]
	\includegraphics[width=0.95\columnwidth]{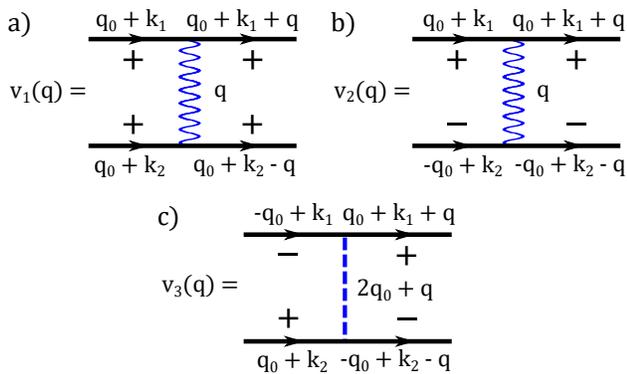}
	\caption{Matrix elements of the electron-electron interaction: (a) the intra-valley matrix element $v_1 (q)$, (b) the direct inter-valley matrix element $v_2 (q)$, (c) the EIV matrix element $v_3 (q)$. Other non-zero matrix elements can be obtained by changing signs of all valley indexes. Here $q \lesssim k_F \ll q_0$, where $k_F$ is the Fermi momentum and $\tau \bm q_0$ is the position of the center of the  valley $\tau$.}
	\label{fig:int}
\end{figure}

The presence of two valleys results in three different scattering channels for interacting electrons: the intra-valley interaction $v_1 (q)$ where interacting electrons are from the same valley, Fig.~\ref{fig:int}(a); the direct inter-valley interaction $v_2 (q)$ where electrons from different valleys scatter without changing the valley, Fig.~\ref{fig:int}(b); the EIV interaction $v_3 (q)$ where electrons exchange their valleys, Fig.~\ref{fig:int}(c).
Matrix elements $v_{1,2} (q)$ are both described by the 
screened Coulomb interaction, see e.g. Ref.~\cite{maslovchubukov}, $v_1(q) = v_2 (q) = 2 \pi e^2 /[\epsilon (q + k_{TF})]$, where $e$ is the electron charge, $\epsilon$ is the dielectric constant, $k_{TF} = 4 / a_B$ is the Thomas-Fermi screening wave vector, $a_B = \epsilon/ m e^2 $ is the effective Bohr radius. In all relevant processes the momentum transfer $q \lesssim k_F$, where $k_F = \sqrt{\pi n}$ is the Fermi momentum, $n$ is the total electron density. At small to moderate densities $k_F a_B \ll 1$, so the term $q a_B$ can be neglected. Neglecting the regular $q$-dependence of the interaction matrix elements does not affect the physics considered here and is done for simplicity.

 The EIV matrix element $v_3 (q)$, Fig.~\ref{fig:int}(c), is short-range due to the large momentum transfer $2 q_0 + q \approx 2 q_0 \gg k_F$, where $2 q_0$ is the distance between two valleys in the first Brillouin zone. Thus, $v_3(q) \approx v_3(0)$. One can estimate $v_3 (0)$ by the bare Coulomb interaction: $v_3 (0) \sim V_C(2 q_0) =  \pi e^2 / \epsilon q_0$. However, this estimate is rather crude as it does not take into account the overlap of Bloch amplitudes corresponding to different valleys and the on-site Hubbard interaction, see Ref.~\cite{roldan} As in TMDs $a_B \sim 1 \,$nm $\sim a_0$, with $a_0$ being the lattice constant, then $v_3(0) \sim v_{1,2}(0)$. For further analysis we introduce the dimensionless interaction matrix elements:
\begin{equation}
v = \nu v_1(0) = \nu v_2 (0) \, > \, u = \nu v_3(0) , \label{vcoul}
\end{equation}
where $\nu = m / 2 \pi$ is the 2D density of states per parabolic band.
In this work we only consider the case  $v > u$. For the opposite case, $v < u$, the system develops a Cooper instability leading to superconducting phases \cite{footnote1}.

\section{Asymmetry between order parameters induced by the EIV scattering}

In this section we demonstrate that the EIV interaction narrows down possible magnetic phases to the ferromagnetic and the paramagnetic ones. This is due to stabilizing effect of the EIV interaction on the valley and spin-valley fluctuations.
In order to examine magnetic instabilities, we lift the assumption that the chemical potential is identical for all bands. Instead, we introduce four different chemical potentials $\mu^\sigma_\tau$, $\tau = \pm 1$, $\sigma = \pm 1 = \uparrow, \downarrow$, see Fig.~\ref{fig:spec}. The SOI defines the quantization axis for the electron spin and thus leads to the following Ising order parameters:
\begin{eqnarray}
M_s = \sum\limits_{\sigma, \tau} \frac{\sigma n^\sigma_\tau}{4 \nu},\ M_v = \sum\limits_{\sigma, \tau} \frac{\tau n^\sigma_\tau}{4 \nu},\ M_\alpha = \sum\limits_{\sigma, \tau} \frac{\sigma \tau n^\sigma_\tau}{4 \nu} ,
\label{mag}
\end{eqnarray}
where $n^\sigma_\tau = - \partial \Omega / \partial \mu^\sigma_\tau$ is the electron density in the band with spin and valley indexes $\sigma$ and $\tau$, $\Omega$ is the grand canonical potential per unit area that depends on $\mu^\sigma_\tau$ and $\alpha$. The introduced order parameters are independent and have dimension of energy. In this description we keep the total electron density $n = \sum_{\sigma, \tau} n^\sigma_\tau$ fixed by the condition:
\begin{equation}
E_F = \frac{1}{4 \nu} \sum\limits_{\sigma, \tau} n^\sigma_\tau = \frac{n}{4 \nu} = const. ,
\label{EF}
\end{equation}
where $E_F$ corresponds to the Fermi energy in the non-interacting case. We emphasize that $E_F$ is not a chemical potential in the system.

In order to find which realizations of the chemical potentials $\mu^\sigma_\tau$ correspond to the ground state of the interacting Fermi gas, we calculate the free energy $F$ per unit area as a function of the magnetizations $M_i$. As the Feynman diagram technique is designed for the grand canonical potential $\Omega$ (per unit area), we first calculate interaction corrections to $\Omega$ and then apply the Legendre transform to find the free energy $F$:
\begin{eqnarray}
F = \Omega + \sum \limits_{\sigma, \tau} \mu^\sigma_\tau n^\sigma_\tau = \Omega - \sum\limits_{\sigma, \tau} \mu^\sigma_\tau \frac{\partial \Omega}{\partial \mu^\sigma_\tau} .
\label{E}
\end{eqnarray}

 The non-interacting part of the grand canonical potential per unit area is the following, see Appendix A:
\begin{eqnarray}
\Omega_0 = - \frac{\nu}{2} \sum\limits_{\sigma, \tau} \left(\mu^\sigma_\tau + \alpha \sigma \tau \right)^2.
\end{eqnarray}
The free energy $F_0$ per unit area of the non-interacting electron system is then given by 
\begin{eqnarray}
F_0 = 2 \nu \left(M_s^2 + M_v^2 + M_\alpha^2 - 2 \alpha M_\alpha \right)  .
\label{E0}
\end{eqnarray}
Here we have suppressed the constant term $2 \nu E_F^2$. The term $- 4 \nu \alpha M_\alpha$  represents the coupling of the magnetization $M_\alpha$ to the SOI $\alpha$ which plays the role of an external field. 
The EIV interaction leads to an asymmetry between the order parameters already in  first order perturbation theory, see Appendix A:
\begin{eqnarray}
 F_1 = - 2 \nu \left[ v_+ M_s^2 + v_- \left(M_v^2 + M_\alpha^2 - 2 \alpha M_\alpha \right) \right],
\label{scb}
\end{eqnarray}
where $v_\pm = v \pm u$. This asymmetry turns out to be very important for the magnetic phase diagram.

\begin{figure}[t]
	\includegraphics[width=0.9\columnwidth]{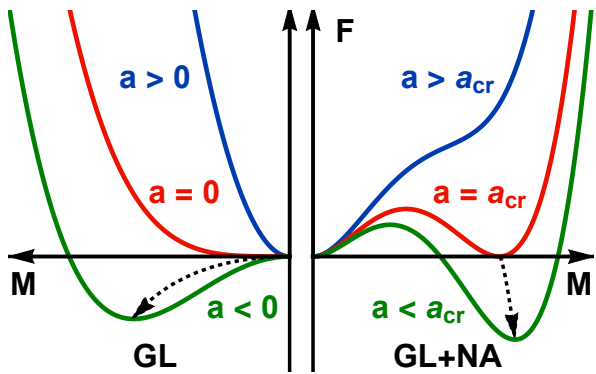}
	\caption{Free energy $F(M)= a M^2 + b M^4 + c |M|^3$ per unit area of the interacting 2DEG as a function of order parameter $M$ with $b > 0$. Left: standard GL free energy with $c=0$. At $a > 0$ the minimum of $F$ is at $M = 0$, at $a < 0$ the minimum is at $M \ne 0$, the critical point $a_{cr} = 0$ corresponds to the second order phase transition. The dashed arrow shows the evolution of minimum at $a < 0$. Right: the GL theory with non-analyticity caused by $c<0$. At $a > a_{cr} = c^2 / 4 b$ the global minimum is at $M = 0$. At $a = a_{cr}$ the ferromagnetic $M \ne 0$ and paramagnetic $M = 0$ minima are degenerate manifesting a first order phase transition. At $a < a_{cr}$ the ferromagnetic minimum is the global one, and its evolution is shown by the dashed arrow. Further evolution of the local paramagnetic minimum $M = 0$ at $a < a_{cr}$ is irrelevant.}
	\label{fig:GL}
\end{figure}

In order to illustrate why the asymmetry between different order parameters is important, we first remind the reader of the Ginzburg-Landau (GL) theory that describes second-order phase transitions. This theory assumes that the free energy is analytic with respect to the order parameters. For example, in case of one order parameter $M$ the GL free energy $F$ reads: $F(M) = a M^2 + b M^4 + \dots$ The parameters $a$, $b$ depend on the interaction, $b > 0$ assures a minimum of $F(M)$ at finite $M$. If the interaction is weak, then $a > 0$ and $F(M)$ has its minimum at $M = 0$ meaning that there is no net magnetization in the ground state. If the interaction is strong such that $a < 0$, then $F(M)$ has its minimum at $|M| = \sqrt{-a / 2 b} \ne 0$ which manifests a ferromagnetic ground state, see Fig.~\ref{fig:GL}. The phase transition occurs at $a = 0$. As there is no magnetization jump at the transition point, this is a second-order phase transition. In our case, $a$ is different for different order parameters: $a_s = 2 \nu (1 - v_+)$ for the spin order parameter $M_s$ and $a_v = a_{\alpha} = 2 \nu (1 - v_-)$ for the valley and spin-valley order parameters, see Eqs.~(\ref{E0}) and  (\ref{scb}). As $v_+  > v_-$ and thus $a_s < a_v = a_{\alpha}$, the GL theory predicts a second-order ferromagnetic phase transition which occurs at $v_+ = 1$ corresponding to $a_s = 0$. Therefore, the critical condition is first satisfied for the ferromagnetic phase. Phases other than ferromagnetic or paramagnetic are not possible due to the energy cost coming from the EIV scattering.

In this paper we make use of the language of magnetizations, see Fig.~\ref{fig:GL}. However, the GL theory can be alternatively understood in terms of the linear response (susceptibility) in the paramagnetic phase. In the above example, the susceptibility $\chi \sim 1 / \frac{\partial^2 F(M)}{\partial M^2}|_{M = 0} \sim 1 / a$ represents the Stoner pole for $a \to 0$ and is only well-defined in the paramagnetic phase where $a > 0$. At the phase transition $a = a_{cr} = 0$ the susceptibility diverges which is the standard signature of a second-order phase transition.

\section{First order phase transition due to the non-analytic terms in the free energy}

In this section we outline the Ginzburg-Landau theory with the negative cubic non-analyticity in case of a single order parameter \cite{belitz}.
The assumption of analyticity of $F(M)$, which underlies the standard GL theory, is not generally true \cite{belitz}.
The non-analytic corrections coming from the infrared electron-hole fluctuations are especially strong in 2D electron systems where the leading non-analyticity is cubic in $M$: $F (M) = a M^2 + b M^4 + c |M|^3 + \dots $, see Refs. \cite{maslovsaha,maslovchubukov,belitz} 
The first two terms in $F (M)$ are exactly the same as in the GL theory. If $c > 0$, then there is no qualitative difference from the GL theory and in this case the theory describes the second-order phase transition at $a_{cr} = 0$. However, if $c < 0$, then the second minimum appears even at $a > 0$. This second minimum becomes the global one at $a < a_{cr} = c^2 / 4 b$ with a finite magnetization jump $|M_{cr}| = -c / 2 b$, see Fig.~\ref{fig:GL}. Note that in this case $a_{cr} > 0$, and the first order phase transition occurs before the Stoner instability of the paramagnetic minimum, corresponding to $a = 0$, develops. In fact, at $a < a_{cr}$ the evolution of the paramagnetic point $M = 0$ is completely irrelevant as the system is settled in the vicinity of the ferromagnetic minimum.
In particular, the Stoner pole in the susceptibility of the paramagnetic phase, $\chi \sim 1 / a$, can  no longer be reached as the phase transition occurs at $a = a_{cr} > 0$. In the ferromagnetic phase the Stoner formula for the susceptibility is no longer valid.
The finite magnetization jump and finite susceptibility of the paramagnetic phase at the critical point demonstrate the first order nature of the phase transition described by the theory with negative cubic non-analyticity, see Fig.~\ref{fig:GL}.

The negative cubic non-analyticity cannot be ignored as it defines the nature of magnetic phase transitions. In our case we have three independent order parameters and the previous theories that considered a single order parameter cannot be applied to our problem. In the next section we calculate the non-analytic cubic correction to the free energy in the leading order in the interaction.

The statement that the EIV scattering narrows down all non-trivial magnetic phases to the ferromagnetic one stays true even when the cubic non-analyticity is taken into account.
Indeed, in the previous section we showed that the smallest $a$ corresponds to the spin order parameter $a_s$, so $a_s$ is always closer than $a_v = a_\alpha$ to the critical value $a_{cr}$ at which the first order phase transition occurs. Therefore, due to the EIV scattering the ferromagnetic instability always wins over the valley and spin-valley instabilities.

The considered mechanism of the first order phase transition relies on the negative cubic non-analyticity in the free energy. It can be independently probed in the paramagnetic phase where this non-analyticity results in an unusual linear magnetic field dependence of the 
spin susceptibility, see Appendix C:
\begin{equation}
	\delta \chi_{zz}  \sim \chi_0 \frac{|\mu_B B_z|}{E_F} ,
	\label{sus}
\end{equation}
where $\chi_0>0$ is the spin susceptibility of the non-interacting 2DEG, $\mu_B$ is the Bohr magneton, and $B_z$ is the perpendicular magnetic field, and the proportionality coefficient is given in the appendix C. 
We emphasize that $\delta \chi_{zz}$ must be positive in order to produce the negative cubic term in the free energy. Similar non-analyticity of the susceptibility in the paramagnetic phase was predicted theoretically for one-valley 2D electron systems \cite{maslovsaha,maslovchubukov,chubukov,zakmaslov,zakQ}. In our case the prefactor is different due to the valley contribution, see Appendix C.

\section{Non-analytic cubic correction to the electron free energy}

Here we calculate the second order Feynman diagrams that contribute to the non-analytic terms in the free energy. The second order diagrams that appear in the RPA series are not included here as they are fully absorbed in the quadratic part of the free energy, see Appendix A.
The diagrams in Fig.~\ref{fig:diag}(a),(b) do not involve the EIV scattering and thus correspond to the one-valley calculations \cite{maslovchubukov,zakmaslov}. In particular, the non-analytic part of the diagram in Fig.~\ref{fig:diag}(b) vanishes. Going beyond this, we now also consider contributions of the processes involving one or two EIV scattering events that are represented by diagrams in Fig.~\ref{fig:diag}(c),(d). The diagram in Fig.~\ref{fig:diag}(d) is especially interesting because its sign is opposite to the diagrams in Fig.~\ref{fig:diag}(a),(c). We outline the calculations of these diagrams in Appendix B.

\begin{figure}[t]
	\includegraphics[width=\columnwidth]{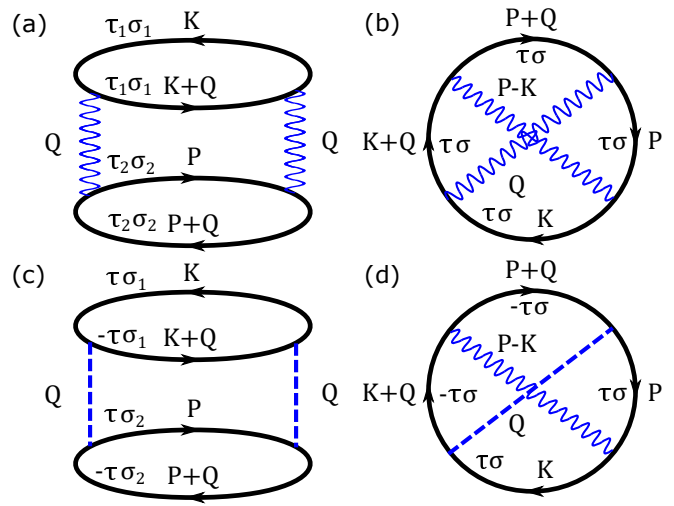}
	\caption{Second order diagrams for the grand canonical potential $\Omega$ beyond the RPA approximation, see Appendix A. Blue wavy (dashed) line corresponds to the $v$ ($u$) component of the effective interaction, index $\tau_i$ ($\sigma_i$) labels the valley (spin). The black solid lines correspond to the bare electron Green function.}
	\label{fig:diag}
\end{figure}

The cubic correction to $\delta \Omega$ given by the diagrams in Fig.~\ref{fig:diag} leads to corresponding correction to the free 
energy $\delta F$, see Eq.~(\ref{E}):
\begin{eqnarray}
&& \delta F = -\frac{2\nu}{3 E_F} \left\{ (v - u)^2 f(M_v, M_\alpha) + u^2 \left(|M_s|^3 + |M_v|^3 \right) \right. \nonumber \\
&&\left. \,\,\,\,\,\,\,\, +\, v^2 \left[f(M_s, M_\alpha) + f(M_s, M_v) \right] \right\} ,
\label{cubic}
\end{eqnarray}
where $f(x, y) = (|x + y|^3 + |x - y|^3) / 2$.
This is the central result of the paper. Contributions from different diagrams can be recognized by the interaction factors $v^2$, $u^2$, and $u v$ corresponding to the diagrams in Fig.~\ref{fig:diag}(a), Fig.~\ref{fig:diag}(c) and Fig.~\ref{fig:diag}(d), respectively (the diagram in Fig.~\ref{fig:diag}(b) does not contribute to the cubic non-analyticity). Notice that Eq.~(\ref{cubic}) contains mixed non-analytic terms like $|M_s \pm M_v|$ which makes it qualitatively different from the case of a single order parameter. Eq.~(\ref{cubic}) is valid under the condition that all order parameters $M_i$ are much smaller than the Fermi energy $E_F$ and corresponds to the limit of zero temperature $T = 0$.
 The temperature effects become important at $T \gtrsim |M^*|$, where $M^*$ is the spontaneous magnetization of the ground state (from one of the order parameters). At this temperature regime the non-analyticity disappears as $|M|^3 \to M^4 / T$. Therefore, one can expect the critical temperature $T_c \sim |M^*|$.

 In order to analyze Eq.~(\ref{cubic}), it is useful to consider three limiting cases when only one of the order parameters is non-zero. If $M_v = M_\alpha = 0$, then $\delta F = - 2 \nu |M_s|^3 (2 v^2 + u^2) / 3 E_F$. For $u = 0$, it reproduces results obtained for the one-valley case of Ref.~\cite{maslovchubukov} If $M_s = M_\alpha = 0$, then $\delta F = -2 \nu |M_v|^3 (2 v^2 + 2 u^2 - 2 u v)/ 3 E_F$. If $M_s = M_v = 0$, then $\delta F = - 2 \nu |M_\alpha|^3 (2 v^2 + u^2 - 2 u v) / 3 E_F$. We notice that the spin term has the largest prefactor $2 v^2 + u^2$. The difference in prefactors is again due to the EIV scattering. Here we point out that the asymmetry mostly arises from the diagram in Fig.~\ref{fig:diag}(d) which reduces prefactors for the valley and spin-valley cubic terms. The sum of $\delta F$ and the quadratic part $F_0+F_1$ [see Eqs.~(\ref{E0}) and  (\ref{scb})] is minimal for the spin term, this allows us to conclude once again that the ferromagnetic instability $M_s \ne 0$, $M_v = M_\alpha = 0$ wins over other possible magnetic instabilities. We stress that the ferromagnetic instability is favored over others because of the EIV scattering. 
 The phase diagram of monolayers of TMDs in absence of the EIV scattering is predicted to be much richer, see Ref.~\cite{donck} This is why in this work we argue that the EIV scattering is an important interaction channel that cannot be neglected.

The Ising ferromagnetic phase was recently observed experimentally  in monolayers of MoS$_2$ \cite{roch}. The reported spin polarization is close to 100\%, i.e. $|M_s| / E_F \approx 1$. Our theory relies on the expansion over small parameters $M_i / E_F$, where $M_i$ are magnetizations defined in  Eq.~(\ref{mag}), and cannot be extended to the case when $M_i /E_F \sim 1$. However, it still allows us to grasp the origin of the observed instability. In particular, we predict the first order phase transition that is in agreement with the large spin magnetization observed in Ref.~\cite{roch} Next, we argue that the ferromagnetic instability is favored due to the EIV scattering. In fact, the limit of total spin magnetization with no valley and spin-valley magnetization corresponds to the case when the EIV scattering contribution to the interaction is maximized. Thus, our theory provides us with the qualitative argument why other phases were not observed.

\section{Conclusions}

In conclusion, we predict the Ising ferromagnetic phase with large spontaneous magnetization in monolayers of TMDs in case of  $E_F \gg \alpha$. The first-order nature of the transition comes from the special role of electron-electron interactions in 2D which gives rise to unusual cubic magnetization terms with a negative sign.  The non-analyticity can be potentially probed by measuring the linear magnetic field dependence of the spin susceptibility in the paramagnetic phase, see Eq.~(\ref{sus}). The Ising order is due to the finite SOI that breaks the $O(3)$ rotational symmetry and defines the quantization axis for spin. The phases with non-zero spontaneous valley polarization cannot develop for $E_F \gg \alpha$ due to the EIV scattering. While our theoretical prediction agrees with recent experiments on MoS$_2$ monolayers based on exciton spectra \cite{roch}, it might be interesting to look for the spontaneous spin polarization directly by e.g. magnetization measurements.

\emph{Acknowledgments.} We would like to thank J. Roch and R. Warburton for very stimulating discussions about their experimental findings which motivated this work. We acknowledge support by the Swiss National Science Foundation and NCCR QSIT as well as the Georg H. Endress foundation. This project received funding from the European Union's Horizon 2020 research and innovation program (ERC Starting Grant, grant agreement No 757725).

\appendix

\section{QUADRATIC PART OF THE FREE ENERGY}

The electron Matsubara Green function corresponding to the Hamiltonian $H$ Eq.~(\ref{ham}) is given by
\begin{equation}
g_\tau^\sigma (K) = \frac{1}{i \omega - \varepsilon (k) + \alpha \sigma \tau + \mu^\sigma_\tau}\,\, ,
\label{green}
\end{equation}
where $K = (\bm k, \omega)$ is the ``relativistic'' notation for the momentum $\bm k = (k_x, k_y)$ and the Matsubara frequency $\omega$, $\varepsilon (k) = k^2 / 2 m$. It is convenient to introduce the notation:
\begin{eqnarray}
\tilde \mu^\sigma_\tau = \mu^\sigma_\tau + \alpha \sigma \tau .
\label{mutilde}
\end{eqnarray}

The non-interacting part of the grand canonical potential (per unit area) is given by the following expression:
\begin{eqnarray}
\Omega_0 = \sum\limits_{\sigma, \tau} \sum_K \ln \left[g^\sigma_\tau (K)\right] ,
\label{O0}
\end{eqnarray} 
where $\sum\limits_K = T \sum\limits_\omega \int d \bm k / (2 \pi)^2$.
Evaluating the sum in Eq.~(\ref{O0}), we get
\begin{equation}
\Omega_0 = - \frac{\nu}{2} \sum \limits_{\sigma, \tau} (\tilde \mu^\sigma_\tau)^2 ,
\label{om0}
\end{equation}
where $\nu = m /2 \pi$ is the 2D density of states.
Using the Legendre transform Eq.~(\ref{E}) and the definition of magnetic order parameters Eq.~(\ref{mag}), we find the non-interacting part of the free energy (per unit area):
\begin{eqnarray}
F_0 = 2 \nu \left(M_s^2 + M_v^2 +M_\alpha^2 - 2 \alpha M_\alpha \right),
\end{eqnarray}
where we suppressed the constant term $2 \nu E_F^2$ which does not depend on the order parameters. The Fermi energy $E_F$ is defined in Eq.~(\ref{EF}).

\begin{figure}[t]
	\includegraphics[width=0.9\columnwidth]{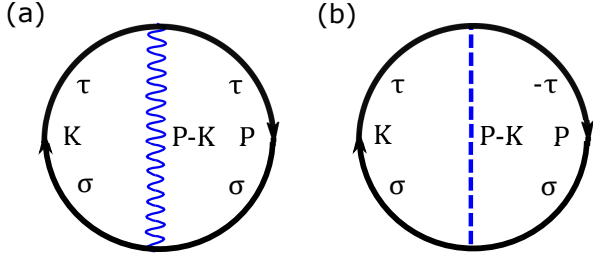}
	\caption{First order interaction correction to the grand canonical potential $\Omega$. Blue wavy (dashed) lines correspond to the $v$ ($u$) components of the effective interaction. The black solid lines correspond to the electron Green function Eq.~(\ref{green}).}
	\label{fig:first}
\end{figure}

\begin{figure}[t]
	\includegraphics[width=0.9\columnwidth]{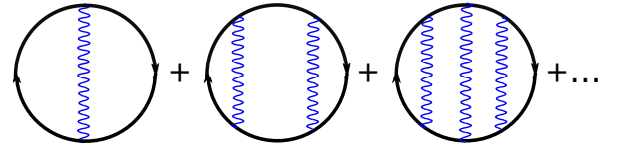}
	\caption{The RPA series for the grand canonical potential $\Omega$. Blue wavy lines represent here both components $v$ and $u$ of the effective interaction for brevity.}
	\label{fig:rpa}
\end{figure}

The first order interaction correction to the grand canonical potential is given by the diagrams in Fig.~\ref{fig:first} and reads explicitly:
\begin{eqnarray}
\!\!\!\!\!\!\!\!\!\!\! \Omega_1 = -\frac{1}{2 \nu} \! \sum\limits_{\sigma, \tau; K, P} \left[v g^\sigma_\tau(K) g^\sigma_\tau (P) 
+ u g^\sigma_\tau(K) g^\sigma_{-\tau} (P) \right] \! ,
\label{om1}
\end{eqnarray}
where the first and second terms correspond to the diagrams in Fig.~\ref{fig:first}(a) and Fig.~\ref{fig:first}(b), respectively.
Here we just have to calculate $\sum_K g^\sigma_\tau (K)$ and find
\begin{eqnarray}
\sum \limits_K g^\sigma_\tau (K) = \nu \tilde \mu^\sigma_\tau .
\end{eqnarray}
Substituting it back to Eq.~(\ref{om1}), we get
\begin{eqnarray}
\Omega_1 = -\frac{\nu}{2} \sum\limits_{\sigma, \tau} \left[v (\tilde \mu^\sigma_\tau)^2 + u \tilde \mu^\sigma_\tau \tilde \mu^\sigma_{-\tau} \right] .
\end{eqnarray}
In order to calculate the correction to the free energy, we have to take into account the renormalization of the order parameters due to $\Omega_1$:
\begin{eqnarray}
&& M_s = (1 + v_+) \sum\limits_{\sigma, \tau} \frac{\sigma}{4} \tilde \mu^\sigma_\tau, \; M_v = (1 + v_-) \sum\limits_{\sigma, \tau} \frac{\tau}{4} \tilde \mu^\sigma_\tau, \nonumber \\
&& M_\alpha = (1 + v_-) \sum\limits_{\sigma, \tau} \frac{\sigma \tau}{4} \tilde \mu^\sigma_\tau ,
\end{eqnarray}
where $v_\pm = v \pm u$. The SOI strength $\alpha$ is also renormalized as $\tilde \alpha = (1 + v_-) \alpha$. In what follows we use the notation $\alpha$ for the renormalized SOI. Applying the Legendre transform, we get the first order interaction correction to the free energy (per unit area)
\begin{eqnarray}
F_1 = -2 \nu \left[v_+ M_s^2 + v_- \left(M_v^2 + M_\alpha^2 - 2 \alpha M_\alpha \right) \right] .
\end{eqnarray}
Together with $F_0$ we get the quadratic part of free energy accounting for the first order interaction correction:
\begin{eqnarray}
F_0 + F_1 = a_s M_s^2 + a_v M_v^2 + a_\alpha (M_\alpha^2 - 2 \alpha M_\alpha) ,
\label{quad}
\end{eqnarray}
where $a_s = 2 \nu (1 - v_+)$, $a_v = a_\alpha = 2 \nu (1 - v_-)$. The higher order diagrams of the random phase approximation (RPA) series [see Fig.~\ref{fig:rpa}] lead to the renormalization of order parameters. In Fig.~\ref{fig:rpa}, wavy lines  represent both components $v$ and $u$ of the interaction for brevity.  It turns out that Eq.~(\ref{quad}) also holds in this case if parameters are replaced by renormalized ones.

\section{NON-ANALYTIC CORRECTION TO THE FREE ENERGY}

Here we outline the calculation of second-order diagrams that first contribute to the non-analytic cubic correction to the grand canonical potential $\Omega$, see Fig.~\ref{fig:diag}, and thus to the free energy $F$. The derivation of these diagrams follows the procedure introduced in Refs. \cite{maslovsaha,maslovchubukov,zakmaslov}
We do not include the second order diagram from the RPA series [see Fig.~\ref{fig:rpa}], as it already has been taken into account. The algebraic representation for all four diagrams in Fig.~\ref{fig:diag} is the following:
\begin{eqnarray}
\!\!\!\!\!\!	\Omega_a \! & = & \! \frac{-v^2}{4 \nu^2} \! \sum \! g^{\sigma_1}_{\tau_1}(K) g^{\sigma_1}_{\tau_1}\!(K+Q) g^{\sigma_2}_{\tau_2}(P) g^{\sigma_2}_{\tau_2}\!(P + Q), \label{xia} \\
\!\!\!\!\!\!	\Omega_b \! & = & \! \frac{v^2}{4 \nu^2} \sum g^{\sigma}_{\tau}(K) g^{\sigma}_{\tau}(K + Q) g^{\sigma}_{\tau}(P+Q) g^{\sigma}_{\tau}(P),
\label{xib} \\
\!\!\!\!\!\!	\Omega_c \! & = & \! \frac{-u^2}{4 \nu^2} \!\! \sum \! g^{\sigma_1}_{\tau}(K) g^{\sigma_1}_{-\tau}\!(K+Q)  g^{\sigma_2}_{\tau}(P) g^{\sigma_2}_{-\tau}\!(P + Q),
\label{xic} \\
\!\!\!\!\!\!	\Omega_d \! & = & \! 2 \frac{u v}{4 \nu^2} \! \! \sum \! g^{\sigma}_{\tau}(K) g^{\sigma}_{-\tau}(K + Q) g^{\sigma}_{-\tau} (P+Q) g^{\sigma}_{\tau}(P), \label{xid}
\end{eqnarray}
where $\nu = m/ 2 \pi$ is the 2D density of states, sums are taken over all spin and valley indexes and over all momenta $K$, $Q$, and $P$. Notice the additional factor of $2$ in $\Omega_d$ which comes from  the fact that $\Omega_d$ is represented by two diagrams: one is shown in Fig.~\ref{fig:diag}(d) and the other is different by swapping dashed and wavy interaction lines.

Next, we perform the summations over $K$ and $P$ merging the Green functions into a particle-hole bubble:
\begin{equation}
\Pi^{\sigma \sigma'}_{\tau \tau'}(Q) = \frac{1}{\nu} \sum_K g^\sigma_\tau(K) g^{\sigma'}_{\tau'}(K + Q) \equiv \mathcal{P}^{\sigma \sigma'}_{\tau \tau'}(Q) - 1,
\end{equation}
where $\mathcal{P}^{\sigma \sigma'}_{\tau \tau'}(Q)$ is the so-called dynamic part of the particle-hole bubble~\cite{maslovsaha,maslovchubukov,zakmaslov}. In fact, we only have to account for the dynamic part of the particle-hole bubbles because the static part is taken into account via the Thomas-Fermi screening of the Coulomb interaction. With this, we get the following expressions for the diagrams:
\begin{eqnarray}
&& \Omega_a = -\frac{v^2}{4} \sum_{\tau_1, \tau_2} \sum_{\sigma_1, \sigma_2} \sum_Q \mathcal{P}_{\tau_1 \tau_1}^{\sigma_1 \sigma_1}(Q) \mathcal{P}_{\tau_2 \tau_2}^{\sigma_2 \sigma_2}(Q) ,
\label{xia2} \\
&& \Omega_b = \frac{v^2}{4} \sum_{\tau, \sigma} \sum_Q [\mathcal{P}^{\sigma \sigma}_{\tau \tau}(Q)]^2,
\label{xib2} \\
&& \Omega_c = -\frac{u^2}{4} \sum_{\tau, \sigma_1, \sigma_2} \sum_Q \mathcal{P}^{\sigma_1 \sigma_1}_{\tau -\tau}(Q) \mathcal{P}^{\sigma_2 \sigma_2}_{\tau -\tau}(Q),
\label{xic2} \\
&& \Omega_d = \frac{u v}{2} \sum_{\tau, \sigma} \sum_Q [\mathcal{P}^{\sigma \sigma}_{\tau -\tau}(Q)]^2.
\label{xid2}
\end{eqnarray}

The non-analyticity comes from the vicinities of two special points of the particle-hole bubble $\mathcal{P}^{\sigma \sigma'}_{\tau \tau'}(Q)$, $Q = (\bm q, \varepsilon)$: (i) the Landau damping at $q = 0$, $\varepsilon = 0$ and (ii) the Kohn anomaly at $q = 2 k_F$, $\varepsilon = 0$, $k_F$ is the Fermi momentum. 
The Landau damping only contributes if at least one of the bubbles $\mathcal{P}^{\sigma \sigma'}_{\tau \tau'} (Q)$ does not have both $\sigma = \sigma'$ and $\tau = \tau'$, see Refs.~\cite{maslovchubukov,zakmaslov} Therefore, the Landau damping gives contributions to $\Omega_c$ and $\Omega_d$ only. For the Kohn anomaly only the difference in chemical potentials of two particle-hole bubbles is important \cite{maslovchubukov}. Therefore, the Kohn anomaly does not contribute to $\Omega_b$ and $\Omega_d$ because the two particle-hole bubbles are identical there. This simple analysis readily shows that neither Landau damping nor Kohn anomaly contribute to $\Omega_b$, meaning that $\Omega_b$ is analytic.

First, we calculate the Landau damping contributions using the following approximation of the particle-hole bubble at small $Q = (\bm q, \varepsilon)$ \cite{maslovsaha,maslovchubukov}:
\begin{eqnarray}
&&\mathcal{P}_{\tau \tau'}^{\sigma \sigma'}(\bm q, \varepsilon) \approx \frac{|\varepsilon|}{\sqrt{\left(\varepsilon - i \Delta^{\sigma \sigma'}_{\tau \tau'}\right)^2 + \left(v_F q\right)^2}}, \label{piss_approx}
\end{eqnarray}
where $v_F = k_F / m$ is the Fermi velocity and 
\begin{equation}
\Delta^{\sigma \sigma'}_{\tau \tau'} = \tilde \mu^\sigma_\tau - \tilde \mu^{\sigma'}_{\tau'} = \mu^\sigma_\tau - \mu^{\sigma'}_{\tau'} + \alpha (\sigma \tau - \sigma' \tau') .
\end{equation} 
Integration over $q$ can be done with the help of the following identity:
\begin{eqnarray}
&&\int\limits_0^\infty \frac{\varepsilon^2 q\, dq/(2 \pi)}{\sqrt{\left(\varepsilon - i \Delta_1\right)^2 + \left(v_F q\right)^2} \sqrt{\left(\varepsilon - i \Delta_2\right)^2 + \left(v_F q\right)^2}}  = \nonumber \\
&&-\frac{\varepsilon^2}{2 \pi v_F^2} \ln \left( 1 - i  \frac{\Delta_1 + \Delta_2 }{2 \varepsilon} \right), \label{int}
\end{eqnarray}
where we have regularized the ultraviolet divergence by subtracting the integral at $\Delta_1 = \Delta_2 = 0$. Therefore, the Landau damping contribution is given by
\begin{eqnarray}
\!\!\!\sum_Q \!\! \vphantom{|}^{L} \mathcal{P}^{\sigma_1 \sigma_2}_{\tau_1 \tau_2}(Q) \mathcal{P}^{\sigma_3 \sigma_4}_{\tau_3 \tau_4}(Q) \!=\! \frac{\nu T^3}{4 E_F}\! \mathcal{F}\!\left(\frac{\Delta^{\sigma_1 \sigma_2}_{\tau_1 \tau_2} + \Delta^{\sigma_3 \sigma_4}_{\tau_3 \tau_4}}{2 T}\right)\!,
\label{Ssum2}
\end{eqnarray}
where the index $L$ of the sum indicates that $Q$ is in the vicinity of the Landau damping point, $\mathcal{F}(z)$ represents the sum over $\varepsilon = 2 \pi n T$, with $T$ being the temperature (Boltzman constant $k_B=1$) and $n$ being an integer:
\begin{equation}
\mathcal{F}(z) = - 2 \sum_\varepsilon \frac{\varepsilon^2}{T^2} \ln \left(1 - i \frac{T z}{\varepsilon} \right).
\label{S}
\end{equation}
This sum is ultraviolet divergent. However, we only need the essential dependence on $z$, meaning that one can ignore the divergent part by rewriting the sum in terms of a contour integral. The integration over the logarithm branch cut readily yields the following representation \cite{maslovchubukov}:
\begin{equation}
\mathcal{F}(z) = \int\limits_0^{z} dx \, x^2 \coth\left(\frac{x}{2}\right).
\label{F}
\end{equation}
Therefore, the Landau damping contribution is given by Eq.~(\ref{Ssum2}) with the function $\mathcal{F}(z)$ given by Eq.~(\ref{F}). At small temperature $T \to 0$ the argument of $\mathcal{F}$ in Eq.~(\ref{Ssum2}) is large and one can use the asymptotic expansion:
\begin{eqnarray}
\mathcal{F}(z) = \frac{|z|^3}{3} + 4 \zeta(3) + O(e^{-|z|}), \quad |z| \gg 1.
\label{expansion}
\end{eqnarray}
We note that Eq.~(\ref{expansion}) establishes the non-analytic behavior of the Landau damping contribution.

In order to calculate the Kohn anomaly contribution, we will use the trick from Ref.~\cite{maslovchubukov} First of all, for the Kohn anomaly only the difference $\delta E_F$ in chemical potential matters:
\begin{equation}
\sum_Q \!\! \vphantom{|}^{K} \mathcal{P}^{\sigma_1 \sigma_2}_{\tau_1 \tau_2}(Q) \mathcal{P}^{\sigma_3 \sigma_4}_{\tau_3 \tau_4}(Q) \! = \! \sum_Q \!\! \vphantom{|}^{K} \mathcal{P}_{\frac{\delta E_F}{2}}(Q) \mathcal{P}_{-\frac{\delta E_F}{2}}(Q),
\label{kohn1}
\end{equation}
where the index $K$ of the sum indicates that $Q$ is in the vicinity of the Kohn anomaly. The difference $\delta E_F$ in chemical potential is the following:
\begin{equation}
\delta E_F = \frac{\tilde \mu_{\tau_1}^{\sigma_1} +\tilde \mu_{\tau_2}^{\sigma_2} -\tilde \mu_{\tau_3}^{\sigma_3} -\tilde \mu_{\tau_4}^{\sigma_4}}{2},
\label{def}
\end{equation}
where $\tilde \mu^\sigma_\tau$ is given by Eq.~(\ref{mutilde}). Next, we notice that Eq.~(\ref{kohn1}) is very similar to the $Q$-integral of $\mathcal{P}_\uparrow \mathcal{P}_\downarrow$ in Ref.~\cite{maslovchubukov} where $\mathcal{P}_\uparrow$ ($\mathcal{P}_\downarrow$) are particle-hole bubbles corresponding to the spin up (down) in a one-valley Fermi gas with Zeeman field. In Ref.~\cite{maslovchubukov} it is shown that one can rewrite the Kohn anomaly contribution of $\mathcal{P}_\uparrow \mathcal{P}_\downarrow$ as the Landau damping contribution of $\mathcal{P}_{\uparrow \downarrow}^2$, where $\mathcal{P}_{\uparrow \downarrow}$ is the particle-hole bubble constructed from up and down components of the Green function. Applying this trick to Eq.~(\ref{kohn1}), we get:
\begin{equation}
\sum_Q \!\! \vphantom{|}^{K} \mathcal{P}^{\sigma_1 \sigma_2}_{\tau_1 \tau_2}(Q) \mathcal{P}^{\sigma_3 \sigma_4}_{\tau_3 \tau_4}(Q) \! = \! \sum_Q \!\! \vphantom{|}^{L} [\mathcal{P}_{\delta E_F}(Q)]^2 ,
\label{kohn2}
\end{equation}
where the sum on the right is taken in the vicinity of Landau damping point.
Applying Eq.~(\ref{Ssum2}), we obtain the Kohn anomaly contribution as
\begin{equation}
\sum_Q \!\! \vphantom{|}^{K} \mathcal{P}^{\sigma_1 \sigma_2}_{\tau_1 \tau_2}(Q) \mathcal{P}^{\sigma_3 \sigma_4}_{\tau_3 \tau_4}(Q) \! = \frac{\nu T^3}{4 E_F} \mathcal{F}\left(\frac{\delta E_F}{T}\right),
\label{kohn3}
\end{equation}
where $\delta E_F$ is given by Eq.~(\ref{def}).

Using Eqs.~(\ref{Ssum2}) and  (\ref{kohn3}) we get the non-analytic parts of the diagrams in Fig.~\ref{fig:diag}:
\begin{eqnarray}
\delta \Omega_a & = & -\frac{\nu v^2 T^3}{16 E_F} \sum_{\tau_i, \sigma_i} \mathcal{F} \left(\frac{\Delta^{\sigma_1 \sigma_2}_{\tau_1 \tau_2}}{T}\right), \quad \delta \Omega_b  =  0, \label{omab}\\
\delta \Omega_c & = &  - \frac{\nu u^2 T^3}{16 E_F} \sum_{\tau, \sigma_i}  \left[ \mathcal{F} \left(\frac{\Delta^{\sigma_1 \sigma_1}_{\tau - \tau} + \Delta^{\sigma_2 \sigma_2}_{\tau - \tau}}{2 T}\right)  \right. \nonumber \\
& + & \left. \mathcal{F} \left(\frac{\Delta^{\sigma_1 \sigma_2}_{\tau \tau} + \Delta^{\sigma_1 \sigma_2}_{-\tau - \tau}}{2 T}\right) \right], \label{omc}\\
\delta \Omega_d & = & \frac{\nu u v T^3}{8 E_F} \sum_{\tau, \sigma} \mathcal{F}\left(\frac{\Delta^{\sigma \sigma}_{\tau - \tau}}{T} \right).
\label{omd}
\end{eqnarray}
As $\delta \Omega_b = 0$, the non-analytic corrections coming from other diagrams can be recognized by the interaction factors $v^2$, $u^2$ and $u v$.
The non-analyticity is especially pronounced at $T = 0$ where the asymptotic expansion Eq.~(\ref{expansion}) of the $\mathcal{F}$-function is valid. Performing the Legendre transform and taking into account the renormalization of order parameters due to the non-analytic correction $\delta \Omega$, we get Eq.~(\ref{cubic}).

\section{MAGNETIC SUSCEPTIBILITY IN THE PARAMAGNETIC PHASE}

In this section, we calculate the non-analytic part of the magnetic field dependence of the magnetic susceptibility in the paramagnetic phase. In this paragraph we only account for the Zeeman coupling and neglect the orbital effects related to the vector-potential. The one-particle Hamiltonian in the paramagnetic phase with the magnetic field $\bm B = (B_x, B_y, B_z)$ is the following
\cite{kormanyos}:
\begin{equation}
H = \frac{k^2}{2 m} - \alpha \sigma_z \tau - \frac{g_s}{2} \mu_B \bm B \cdot \sig - \frac{g_v}{2} \mu_B B_z \tau ,
\label{ham2}
\end{equation}
where $\sig = (\sigma_x, \sigma_y, \sigma_z)$ are the spin Pauli matrices, $g_s$ and $g_v$ are the spin and valley $g$-factors, $\mu_B$ is the Bohr magneton.

We define the magnetic susceptibility as follows:
\begin{eqnarray}
\chi_{i j} = - \frac{\partial^2 \Omega}{\partial B_i \partial B_j} .
\end{eqnarray}
Here we are interested in $\chi_{zz}$ and $\chi_{xx} = \chi_{yy}$. 
Note that for the susceptibility $\chi_{zz}$ defined this way the formulation in terms of the linear response functions should be done using $\langle (g_s \sigma_z + g_v \tau) (g_s \sigma_z + g_v \tau) \rangle$ correlation function instead of the standard spin-spin correlation function $\langle g_s \sigma_z g_s \sigma_z \rangle$. This is because the magnetic field $B_z$ couples both to the spin and the valley, see Eq.~(\ref{ham2}).

If the magnetic field is perpendicular to the 2DEG plane, then we consider eigenstates of $\sigma_z = \sigma = \pm 1$. For the non-analytic correction to the grand canonical potential $\Omega$ we can readily use Eqs.~(\ref{omab})--(\ref{omd}) thanks to the correspondence
\begin{eqnarray}
\tilde \mu^\sigma_\tau = \alpha \sigma \tau + \frac{g_s}{2} \mu_B B_z \sigma + \frac{g_v}{2} \mu_B B_z \tau .
\label{corr}
\end{eqnarray}
As the SOI in TMDs is not smaller than a few meV,  we always assume here that the magnetic field splitting is always smaller than the SOI.
The corresponding non-analytic correction $\delta \chi_{zz}$ to the susceptibility at zero temperature is then linear in the magnetic field:
\begin{align}
& \frac{\delta \chi_{zz}}{\chi_0} = \frac{|\mu_B B_z|}{2 E_F} \frac{v^2 f(g_s, g_v) + u^2 (|g_s|^3 + |g_v|^3)}{g_s^2 + g_v^2} \nonumber \\
& \hspace{60pt}+ \frac{|\alpha|}{E_F} \frac{v^2 g_s^2 + (v - u)^2 g_u^2}{g_s^2 + g_v^2} ,
\end{align}
where $f (x, y) = (|x + y|^3 + |x - y|^3) / 2$, and $\chi_0 = \nu \mu_B^2 (g_s^2 + g_v^2)$ is the magnetic susceptibility of the non-interacting 2DEG. The linear non-analyticity in $\chi_{zz}$ is due to the cubic non-analyticity in the grand canonical potential. The same non-analyticity results in the first order phase transition. Therefore, measuring the unusual linear dependence of the magnetic susceptibility on the magnetic field allows to probe directly the non-analytic cubic correction to the grand canonical potential $\Omega$.

In case of an in-plane magnetic field the quantization axes in different valleys do not coincide. This leads to non-trivial spin traces in the diagrams in Fig.~\ref{fig:diag}. Apart from this complication, the rest is the same. For example, instead of Eqs.~(\ref{xia2})--(\ref{xid2}), we get similar expressions:
\begin{eqnarray}
&& \Omega_a = -\frac{v^2}{4} \sum_{\tau_1, \tau_2} \sum_{\sigma_1, \sigma_2} \sum_Q \mathcal{P}_{\tau_1 \tau_1}^{\sigma_1 \sigma_1}(Q) \mathcal{P}_{\tau_2 \tau_2}^{\sigma_2 \sigma_2}(Q) ,
\label{xia22} \\
&& \Omega_b = \frac{v^2}{4} \sum_{\sigma, \tau} \sum_Q \mathcal{P}^{\sigma \sigma}_{\tau \tau}(Q)^2,
\label{xib22} \\
&& \Omega_c = -\frac{u^2}{4} \sum_{\tau, \sigma_i} \! \mathcal{B}^{\sigma_1 \sigma_2} \mathcal{B}^{\sigma_3 \sigma_4} \! \sum_Q \! \mathcal{P}^{\sigma_1 \sigma_2}_{\tau -\tau}(Q) \mathcal{P}^{\sigma_3 \sigma_4}_{\tau -\tau}(Q),
\label{xic22} \\
&& \Omega_d = \frac{u v}{2} \sum_{\tau, \sigma_1, \sigma_2} \mathcal{B}^{\sigma_1 \sigma_2} \sum_Q \mathcal{P}^{\sigma_1 \sigma_2}_{\tau -\tau}(Q)^2 ,
\label{xid22}
\end{eqnarray}
where the spin tensors $\mathcal{B}^{\sigma_1 \sigma_2}$ are due to the non-collinear spin quantization axes in different valleys and given by
\begin{eqnarray}
\mathcal{B}^{\sigma_1 \sigma_2} = \frac{1}{2} \left(1 + \sigma_1 \sigma_2 \frac{\Del_{+} \cdot \Del_{-}}{\Delta_{+} \Delta_{-}}\right) .
\label{B}
\end{eqnarray}
Here, $\Delta_\tau = |\Del_\tau|$ is the spin gap in valley $\tau$ with
\begin{equation}
\Del_\tau = \left(\frac{g_s}{2} \mu_B B_x, \frac{g_s}{2} \mu_B B_y, \frac{g_s}{2} \mu_B B_z + \alpha \tau \right) .
\label{deltau}
\end{equation}

Using Eqs.~(\ref{Ssum2}) and (\ref{kohn3}) we get the non-analytic parts of the diagrams defined in  Eqs.~(\ref{xia22})--(\ref{xid22}),
\begin{widetext}
	\begin{eqnarray}
\!\!\!\!\!\!\!\!\!\!\!	\delta \Omega_a \! & = & \! -\frac{\nu v^2 T^3}{8 E_F} \left[ \mathcal{F}\left(\frac{2 \Delta_+}{T}\right) + \mathcal{F}\left(\frac{2 \Delta_-}{T}\right) +  \sum_{\sigma_1, \sigma_2} \mathcal{F}\left(\frac{\Delta_+ + \sigma_1 \Delta_- + \sigma_2 g_v B_z}{T}\right) \right], \quad \delta \Omega_b  =  0, \label{omab2}\\
\!\!\!\!\!\!\!\!\!\!\!	\delta \Omega_c \! & = & \! - \! \frac{\nu u^2 T^3}{8 E_F} \!\! \sum_{\sigma_i} \! \mathcal{B}^{\sigma_1 \sigma_2} \mathcal{B}^{\sigma_3 \sigma_4}\! \left[ \mathcal{F}\!\left(\! \frac{(\sigma_1 + \sigma_3) \Delta_+ + (\sigma_2 + \sigma_4) \Delta_-}{2T}\right) \! + \! \mathcal{F}\!\left(\frac{(\sigma_1 + \sigma_3) \Delta_+ - (\sigma_2 + \sigma_4) \Delta_- + 2 g_v B_z}{2T} \right) \!\right]\!, \label{omc2}\\
\!\!\!\!\!\!\!\!\!\!\!	\delta \Omega_d \! & = & \! \frac{\nu u v T^3}{4 E_F} \sum_{\sigma_1, \sigma_2} \mathcal{B}^{\sigma_1 \sigma_2} \mathcal{F}\left(\frac{\sigma_1 \Delta_+ - \sigma_2 \Delta_- + g_v B_z}{T} \right).
	\label{omd2}
	\end{eqnarray}
\end{widetext}
In case of $B_x = B_y = 0$, Eqs.~(\ref{omab2})--(\ref{omd2}) coincide with Eqs.~(\ref{omab})--(\ref{omd}) upon the correspondence outlined in Eq.~(\ref{corr}).

In case if $B_z = 0$ and $T = 0$, Eqs.~(\ref{omab2})--(\ref{omd2}) can be further simplified:
\begin{eqnarray}
\!\!\! \!\!\!\!\! \!\!\delta \Omega \! = \! -\!\frac{4 \nu |\Delta|^3}{3 E_F} \!\! \!\left(\!\! v^2 \! + \! \frac{u^2}{2} \!\! \left(\! 1 \! - \! \frac{(g_s \mu_B B_\parallel)^2 \alpha^2}{4 \Delta^4} \! \right)\! \! - \! u v \frac{\alpha^2}{\Delta^2} \!\right)\!\!,
\label{inp}
\end{eqnarray}
where $\delta \Omega$ is the total non-analytic correction, $B_\parallel = \sqrt{B_x^2  + B_y^2}$ is the in-plane magnetic field, $\Delta = \sqrt{\alpha^2 + (g_s \mu_B B_\parallel/2)^2}$. It follows from Eq.~(\ref{inp}) that there is no linear dependence of the in-plane susceptibility on the magnetic field in case $g_s \mu_B B_\parallel \ll \alpha$ as the SOI vector $(0, 0, \alpha)$ is orthogonal to the in-plane magnetic field. Therefore, only the $\chi_{zz}$ component of the magnetic susceptibility is useful to detect the cubic non-analyticity.


\begin{thebibliography}{10}

\bibitem{roch} J.~G.~Roch, G.~Froehlicher, N.~Leisgang, P.~Makk, K.~Watanabe, T.~Taniguchi, and R.~J.~Warburton, Nat. Nanotechnol.
\textbf{14}, 432--436 (2019).

\bibitem{kelly} M.~J.~Kelly and L.~M.~Falicov, Phys. Rev. B \textbf{15}, 1974 (1977).
\bibitem{bloss} W.~L.~Bloss, L.~J.~Sham, and V.~Vinter, Phys. Rev. Lett. \textbf{43}, 1529 (1979).
	
\bibitem{wang}	Q. H. Wang, K. Kalantar-Zadeh, A. Kis, J. N. Coleman, and M. S. Strano,	Nat. Nanotechnol. \textbf{7}, 699 (2012).
	
	
\bibitem{zhu} Z.~Y.~Zhu, Y.~C.~Cheng, and U.~Schwingenschl$\rm{\ddot o}$gl, Phys. Rev. B \textbf{84}, 153402 (2011).
\bibitem{xiao} D.~Xiao, G.-B.~Liu, W.~Feng, X.~Xu, and W.~Yao, Phys. Rev. Lett. \textbf{108}, 196802 (2012).
\bibitem{chei} T.~Cheiwchanchamnangij and W.~R.~L.~Lambrecht, Phys. Rev. B \textbf{85}, 205302 (2012).
\bibitem{kormanyos2013}  A. Kormanyos, V. Zolyomi, N. D. Drummond, P. Rakyta, G. Burkard, and V. I. Falko, Phys. Rev. B {\textbf 88}, 045416 (2013).
	
\bibitem{klinovaja} J. Klinovaja and D. Loss,  Phys. Rev. B \textbf{88}, 075404 (2013).
\bibitem{review_1} X. Xu, W. Yao, D. Xiao, and T. F. Heinz, Nature Phys. {\bf 10}, 343 (2014).
\bibitem{pisoni} R.~Pisoni, A.~Kormanyos, M.~Brooks, Z.~Lei, P. Back, M.~Eich, H.~Overweg, Y.~Lee, P.~Rickhaus, K.~Watanabe, T.~Taniguchi, A.~Imamoglu, G.~Burkard, T.~Ihn, and K.~Ensslin, Phys. Rev. Lett. \textbf{121}, 247701 (2018).
\bibitem{kormanyos} A.~Kormanyos, V.~Zolyomi, N.~D.~Drummond, and G.~Burkard, Phys. Rev. X \textbf{4}, 011034 (2014).
	
	
\bibitem{cheng} Y.~C.~Cheng, Q.~Y.~Zhang, and U.~Schwingenschl$\rm{\ddot o}$gl, Phys. Rev. B \textbf{89}, 155429 (2014).
\bibitem{aivazian} G.~Aivazian, Z.~Gong, A.~M.~Jones, R.-L.~Chu, J.~Yan, D.~G.~Mandrus, C.~Zhang, D.~Cobden, W.~Yao, and X.~Xu, Nature Phys. \textbf{11}, 148 (2015).
	
\bibitem{zeng} H.~Zeng, J.~Dai, W.~Yao, D.~Xiao, and X.~Cui, Nat. Nanotechnol. \textbf{7}, 490 (2012).
\bibitem{mak} K.~F.~Mak, K.~He, J.~Shan, and T.~F.~Heinz, Nat. Nanotechnol. \textbf{7}, 494 (2012).
	
	
\bibitem{donck} M. Van der Donck and F. M. Peeters, Phys. Rev. B \textbf{98}, 115432 (2018).
\bibitem{mukherjee} D.~K.~Mukherjee, A.~Kundu, and H.~A.~Fertig, Phys. Rev. B \textbf{98}, 184413 (2018).
\bibitem{braz} J.~E.~H.~Braz, B.~Amorim, and E.~V.~Castro, Phys. Rev. B \textbf{98}, 161406(R) (2018).
	
\bibitem{scrace} T.~Scrace, Y.~Tsai, B.~Barman, L.~Schweidenback, A.~Petrou, G.~Kioseoglou, I.~Ozfidan, M.~Korkusinski, and P.~Hawrylak, Nat. Nanotechnol. \textbf{10}, 603 (2015).
	
		
	
\bibitem{maslovsaha} D.~L.~Maslov, A.~V.~Chubukov, and R.~Saha, Phys. Rev. B \textbf{74}, 220402(R) (2006).
\bibitem{maslovchubukov} D.~L.~Maslov and A.~V.~Chubukov, Phys. Rev. B \textbf{79}, 075112 (2009).
\bibitem{chubukov} A.~V.~Chubukov and D.~L.~Maslov, Phys. Rev. B \textbf{68}, 155113 (2003).
\bibitem{zakmaslov} R.~A.~Zak, D.~L.~Maslov, and D.~Loss, Phys. Rev. B \textbf{82}, 115415 (2010).
\bibitem{zakQ} R.~A.~Zak, D.~L.Maslov, and D.~Loss, Phys. Rev. B \textbf{85}, 115424 (2012).
	
	
	

\bibitem{roldan} R.~Roldan, E.~Cappelluti, and F.~Guinea, Phys. Rev. B \textbf{88}, 054515 (2013).


\bibitem{footnote1}	The inter-pocket Cooper pairing due to large EIV scattering $u > v$ is believed to be the main mechanism of high-T$_c$ superconductivity in iron pnictides \cite{vavilov}. Recent experiments in TMDs at high densities also indicate a superconducting dome \cite{ye,lu,costanzo} which might be due to the inter-valley $u > v$ pairing \cite{roldan}.
	
\bibitem{vavilov} A.~V.~Chubukov, M.~G.~Vavilov, and A.~B.~Vorontsov, Phys. Rev. B \textbf{80}, 140515(R) (2009).
\bibitem{ye} J.~T.~Ye, Y.~J.~Zhang, R.~Akashi, M.~S.~Bahramy, R.~Arita, Y.~Iwasa, Science \textbf{338}, 1193 (2012).
\bibitem{lu} J.~M.~Lu, O.~Zheliuk, I.~Leermakers, N.~F.~Q.~Yuan, U.~Zeitler, K.~T.~Law, and J.~T.~Ye, Science \textbf{350}, 1353 (2015).
\bibitem{costanzo} D. Costanzo, H. Zhang, B. A. Reddy, H. Berger, and A. F. Morpurgo, Nat. Nanotechnol. \textbf{13}, 483 (2018).

\bibitem{belitz} D.~Belitz, T.~R.~Kirkpatrick, and T.~Vojta,
Rev. Mod. Phys. \textbf{77}, 579 (2005).
	
\end{thebibliography}
\end{document}